\documentstyle[amstex,amssymb,12pt,epsfig]{article}

\begin{document}

\bigskip

\font\ninerm= cmr9

%

%

%

%

%

\pagestyle{empty}

\begin{flushright}
hep-th/0209254 \\
$~$ \\
September 2002
\end{flushright}

\vskip0.5 cm

\begin{center}
%

{\large {\bf On the IR/UV mixing and experimental limits\\[0pt]
on the parameters of canonical noncommutative spacetimes}}
\end{center}

\vskip1.5 cm

\begin{center}
{\footnotesize {\bf Giovanni~AMELINO-CAMELIA}, {\bf Gianluca~MANDANICI} and
{\bf Kensuke~YOSHIDA}}\\[0pt]

{\it Dipart.~Fisica, Univ.~Roma ``La Sapienza'', P.le Moro 2, 00185 Roma,
Italy}
\end{center}

\vspace{1cm}

\begin{center}
{\bf ABSTRACT}
\end{center}

{\leftskip=0.6in \rightskip=0.6in We investigate some issues that are
relevant for the derivation of experimental limits on the parameters of
canonical noncommutative spacetimes. By analyzing a simple Wess-Zumino-type
model in canonical noncommutative spacetime with soft supersymmetry breaking
we explore the implications of ultraviolet supersymmetry on low-energy
phenomenology. The fact that new physics in the ultraviolet can modify
low-energy predictions affects significantly the derivation of limits on the
noncommutativity parameters based on low-energy data. These are, in an
appropriate sense here discussed, ``conditional limits''.
We also find that some standard techniques for an effective
low-energy description of theories with non-locality at short distance
scales are only applicable in a regime where theories in canonical
noncommutative spacetime lack any predictivity, because of the
strong sensitivity to unknown UV physics.
It appears useful
to combine high-energy data, from astrophysics, with the more readily
available low-energy data.}

\newpage

%
\baselineskip12pt plus .5pt minus .5pt

\pagenumbering{arabic}

\pagestyle{plain}

\section{Introduction}
Recently, there has been strong interest (see, {\it e.g.}, Refs.~\cite
{con,mrs,MST}) in quantum fields theories constructed on
canonical\footnote{Interest in the $\kappa $-Minkowski~\cite{kappa}
Lie-algebra noncommutative spacetime, $\left[x_{m},t\right]=i\lambda x_{m}$,
$\left[ x_{m},x_{l} \right] =0$, has also grown recently,
especially because of its possible
role in the new relativistic theories with two observer-independent
scales~\cite{dsr1,jurekDSRnew}. We here focus exclusively on canonical
noncommutative spacetimes (\ref{cns}).} noncommutative spacetime:
\begin{equation}
\left[ x_{\mu },x_{\nu }\right] =i\theta _{\mu \nu }~.  \label{cns}
\end{equation}
This recent interest is mostly due to the possible use of these spacetimes
in effective-theory descriptions of string theory in presence of an external
background field, in which case $\theta _{\mu \nu }$ reflects the properties
of the background. Previously the same algebraic relations provided the
basis~\cite{dop,filk} for an approach to the fundamental description of
spacetime physics.

A key characteristic of field theories on canonical spacetimes, which
originates from the commutation rules, is nonlocality. At least in the case
of space/space noncommutativity ($\theta _{0i}=0$), to which we limit our
analysis for simplicity\footnote{The case of space/time
noncommutativity ($\theta _{0i}\neq 0$) is not
necessarily void of interest~\cite{stgood}, but it is more delicate,
especially in light of possible concerns for unitarity. Since our analysis
is not focusing on this point we will simply assume that $\theta _{0i}=0$.},
this nonlocality is still tractable although it induces a characteristic
mixing of the ultraviolet and infrared sectors of the theory. This IR/UV
mixing has wide implications, including the possible emergence of infrared
(zero-momentum) poles in the one-loop two-point functions. In particular one
finds a quadratic pole for some integer-spin particles in non-SUSY theories~
\cite{mrs}, while in SUSY theories the poles, if at all present, are
logarithmic~\cite{MST,giro,KT}. It is noteworthy that these infrared
singularities are introduced by loop corrections and originate from the
ultraviolet part of the loop integration: at tree level the two-point
functions are unmodified, but loop corrections involve the interaction
vertices, which are modified already at tree level.

There has been considerable work attempting to set limits on the
noncommutativity parameters $\theta$ by exploiting the modifications of the
interaction
vertices~\cite{propphen,propvert1,propvert2,propvert3}
and the modifications of the  dressed/full propagators~\cite{gl}.
Most of these analyses rely on our readily
available low-energy data. The comparison between theoretical predictions
and experimental data is usually done using a standard strategy (the methods
of analysis which have served us well in the study of conventional theories
in commutative spacetime). We are here mainly interested in understanding
whether one should take into account some of the implications of the IR/UV
mixing also at the level of the techniques by which one compares theoretical
predictions with data. In Ref.~\cite{gl} it was argued that the way in which
low-energy data can be used to constrain the noncommutativity parameters is
affected by the IR/UV mixing. These limits on the entries of the $\theta $
matrix might not have the usual interpretation: they could be seen only as
``conditional limits'', conditioned by the assumption that no contributions
relevant for the analysis are induced by the ultraviolet. The study we
report here is relevant for this delicate issue. By analyzing a simple
noncommutative Wess-Zumino-type model, with soft supersymmetry breaking, we
explore the implications of ultraviolet supersymmetry on low-energy
phenomenology. Based on this analysis, and on the intuition it provides
about other possible features of ultraviolet physics, we provide a
characterization of low-energy limits on the noncommutativity parameters.
Our analysis provides additional encouragement for combining, as proposed in
Ref.~\cite{gl}, high-energy data, from astrophysics, with the more readily
available low-energy data.

\section{Preliminaries on the IR/UV mixing}
The construction of quantum field theories in canonical noncommutative
spacetime is usually obtained by means of the Weyl map that acts between the
functions on noncommutative ${\cal R}^{4}$ and the functions on commutative $%
{\cal R}^{4}$. The Weyl map associates to the product of two functions on
noncommutative ${\cal R}^{4}$ the ``$\star $'' (Moyal) product
\begin{equation}
f(x)\star g(x):=\left. e^{\frac{i}{2}\theta ^{\mu \nu }\partial _{\mu
}^{y}\partial _{\nu }^{z}}f(y)g(z)\right| _{y=z=x}~,
\end{equation}
which is noncommutative, but associative. As mentioned, at tree level
the $\star$ product induces a modification of the interaction vertices, which
acquire characteristic $\theta $- and momentum-dependent phases, while the
tree-level propagator is unaffected. At one-loop level the modified vertices
generate $\theta $-dependent corrections to the propagator. Let us revisit
briefly, in the illustrative example of the ``$\lambda \Phi ^{4}$''
scalar-boson field theory, the IR/UV mixing that affects
these $\theta$-dependent corrections to the propagator.
Adopting the standard strategy
of distinguishing between planar and nonplanar
diagrams~\cite{mrs,MST,gacluisamichele}, one finds a planar tadpole
contribution characterized by integrals of the form:
\begin{equation}
\int_{0}^{\Lambda }dk\frac{k^{3}}{k^{2}+m^{2}}=\frac{1}{2}\Lambda ^{2}
-\frac{1}{2}m^{2}\ln (1+\frac{\Lambda ^{2}}{m^{2}}).  \label{C}
\end{equation}
and a corresponding nonplanar tadpole contribution of the type:
\begin{equation}
\int_{0}^{\Lambda }dk\cos (\frac{1}{2}k\widetilde{p})\frac{k^{3}}{k^{2}
+m^{2}} ~,  \label{NC}
\end{equation}
where we introduced a momentum cutoff $\Lambda $ and the standard notation $%
\widetilde{p}_{\mu }\equiv \theta _{\mu \nu }p_{\nu }$.

As well known, the integral (\ref{C}) is cut off by $\Lambda $ while its
nonplanar counterpart (\ref{NC}), is cut off by the smaller between $\Lambda
$ and $|\widetilde{p}|^{-1}$. In fact, for $\Lambda \ll |\widetilde{p}|^{-1}$
the $\theta $-dependent phase in (\ref{NC}) is insignificant, while for $%
\Lambda \gg |\widetilde{p}|^{-1}$ the integrand of (\ref{NC}) oscillates
rapidly in the region where the integration momentum $k$ is such that $k\gg |%
\widetilde{p}|^{-1}$:
\begin{equation}
\int_{0}^{\Lambda }dk\cos (\frac{1}{2}k\widetilde{p}) \frac{k^{3}}{%
k^{2}+m^{2}}\simeq \frac{1}{2}\left( \frac{2}{\left| \widetilde{p}\right| }%
\right) ^{2}- \frac{1}{2}m^{2}\ln (1+\frac{\left( \frac{2}{\left| \widetilde{%
p}\right| } \right) ^{2}}{m^{2}}) ~,  \label{formula}
\end{equation}

The planar diagram, which is also present in the corresponding
commutative-spacetime theory, diverges in the usual $\Lambda \rightarrow
\infty $ limit. Instead the $\Lambda \rightarrow \infty $ limit of the
nonplanar diagram is still finite as long as $\widetilde{p}\neq 0$. The
divergence emerges only in the $\widetilde{p}\rightarrow 0$ (infrared)
limit. Just like the UV portion of the loop integration introduces the $%
\Lambda $ dependence of the planar diagram (\ref{C}), it is the UV portion
of the loop integration that introduces the dependence on $\frac{1}{\left|%
\widetilde{p}\right| }$, and the associated infrared singularity, of the
nonplanar diagrams (\ref{NC}). This is a key aspect of the IR/UV mixing.

\section{Effects of UV SUSY on IR physics}
In this section we analyze a mass deformed Wess-Zumino model in canonical
noncommutative spacetime. We emphasize the role that the UV scale of SUSY
restoration plays in the IR sector of the model, and we also provide some
more general remarks on the IR/UV mixing. This analysis will provide
material for one of the points we raise in the later part of the paper,
which concerns the nature of the bounds that can be set on the
noncommutativity parameters using low-energy data.

\subsection{A model with SUSY restoration in the UV}
For definiteness, we present our observations, which have rather wide
applicability, in the specific context of a mass deformed Wess-Zumino model,
with action
\begin{align}
S_{dwz}& =S_{0}+S_{m}+S_{g}, \\
S_{0}& =\int dx^{4}\left\{ \frac{1}{2}\partial _{\mu }\varphi _{1}\partial
^{\mu }\varphi _{1}+\frac{1}{2}\partial _{\mu }\varphi _{2}\partial ^{\mu
}\varphi _{2}+\frac{1}{2}\overline{\psi }i\setbox0=\hbox{$p$}\dimen0=\wd0%
\setbox1=\hbox{/}\dimen1=\wd1\ifdim\dimen0>\dimen1%
\rlap{\hbox to
\dimen0{\hfil/\hfil}}\partial \else\rlap{\hbox
to \dimen1{\hfil$p$\hfil}}/\fi\psi \right\} , \\
S_{m}& =\int dx^{4}\left\{ \frac{1}{2}F^{2}+\frac{1}{2}G^{2}+m_{s}F\varphi
_{1}+m_{s}G\varphi _{2}-\frac{1}{2}m_{f}\overline{\psi }\psi \right\} , \\
S_{g}& =\int dx^{4}g\left\{ F\star \varphi _{1}\star \varphi _{1}-F\star
\varphi _{2}\star \varphi _{2}+G\star \varphi _{1}\star \varphi _{2}+\right.
\\
& \left. \text{ \ \ \ \ \ \ \ \ \ \ }+G\star \varphi _{2}\star \varphi_{1}-
\overline{\psi }\star \psi \star \varphi_{1}-\overline{\psi }\star i\gamma
^{5}\psi \star \varphi_{2}\right\} ~.  \nonumber
\end{align}
$\varphi_{1}$ and $\varphi_{2}$ are bosonic/scalar degrees of freedom, while
$\psi$ denotes fermionic spin-$1/2$ degrees of freedom. $F$ and $G$ are
auxiliary fields. The model is exactly supersymmetric (SUSY)
if $m_{s} = m_{f}$.
We consider the case $m_{s} < m_{f}$ in which supersymmetry is
only ``restored" in the ultraviolet (UV), where both $m_{s}$ and $m_{f}$ are
negligible with respect to the high momenta involved.

The free propagators are not modified by canonical noncommutativity:
\begin{align}
\Delta _{m_{s}}(p)& \equiv \Delta _{\varphi _{1}\varphi _{1}}(p)=\Delta
_{\varphi _{2}\varphi _{2}}(p)=\frac{i}{p^{2}-m_{s}^{2}+i\varepsilon }%
~,~~~~~~~~~~~\Delta _{FF}(p)=\Delta _{GG}(p)=p^{2}\Delta _{\varphi
_{1}\varphi _{1}}(p), \\
\Delta _{F\varphi _{1}}(p)& =\Delta _{\varphi _{1}F}(p)=\Delta _{\varphi
_{2}G}(p)=\Delta _{G\varphi _{2}}(p)=-m_{s}\Delta _{\varphi _{1}\varphi
_{1}}(p)~,~~~~~~~~~~~S(p)=\frac{i}{\setbox0=\hbox{$p$}\dimen0=\wd0\setbox1=%
\hbox{/}\dimen1=\wd1\ifdim\dimen0>\dimen1\rlap{\hbox to \dimen0{\hfil/\hfil}}%
p\else\rlap{\hbox
to \dimen1{\hfil$p$\hfil}}/\fi-m_{f}}~.  \nonumber
\end{align}
The vertices acquire the familiar $\theta $-dependent phases:
\begin{align}
V_{[\overline{\psi }\psi \varphi _{1}]}=-ig\cos (p_{1}\widetilde{p}_{2})&
~,~~~V_{[\overline{\psi }\psi \varphi _{2}]}=-i\gamma ^{5}g\cos (p_{1}%
\widetilde{p}_{2})~,~~~ \\
V_{[F\varphi _{1}\varphi _{1}]}=ig\cos (p_{1}\widetilde{p}%
_{2})~,~~~V_{[F\varphi _{2}\varphi _{1}]}=& -ig\cos (p_{1}\widetilde{p}%
_{2})~,~~~V_{[G\varphi _{1}\varphi _{2}]}=2ig\cos (p_{1}\widetilde{p}_{2})~.
\nonumber
\end{align}
[Notice that, taking into account momentum conservation at vertices, the
momenta $p_{1}$and $p_{2}$ can be attributed equivalently to any of the
three particles involved in each of the vertices.]

\subsection{Self-energies and IR singularities}
Self-energies will play a key role in our observations. Using the NC Feynman
rules the self-energies for fermions and scalars can be evaluated
straightforwardly. The one loop self-energy of the scalar field receives
contributions from five Feynman diagrams, leading to the result
\begin{align}
-i\Sigma _{1loop}(p)& =-g^{2}\int \frac{d^{4}k}{\left( 2\pi \right) ^{4}}%
\left\{ \left( 8k^{2}+8m_{s}^{2}\right) \Delta _{m_{s}}(p)\Delta
_{m_{s}}(p+k)+\right. \\
& \left. -\left( 8k^{2}+8m_{f}^{2}+8p{{\cdot} }k\right) \Delta
_{m_{f}}(p)\Delta _{m_{f}}(p+k)\right\} \cos ^{2}(k\widetilde{p}).
\end{align}

This expression can be seen as the sum of three terms, and each of these
terms is the sum of a planar and of a nonplanar part: $-i\Sigma_{1loop}(p) =
I_{1}^{P}(p) +I_{1}^{NP}(p) +I_{2}^{P}(p)+I_{2}^{NP}(p)
+I_{3}^{P}(p)+I_{3}^{NP}(p)$ with

\medskip \noindent $I_{1}^{P}(p)+I_{1}^{NP}(p) \equiv \frac{1}{2}g^{2}\int
\frac{dk^{4}}{\left( 2\pi \right) ^{4}}\frac{8k^{2}+8m_{s}^{2}}{\left(
k^{2}-m_{s}^{2}\right) \left( (k+p)^{2}-m_{s}^{2}\right) }+\frac{1}{2}%
g^{2}\int \frac{dk^{4}}{\left( 2\pi \right) ^{4}}\cos (2p\widetilde{k})\frac{%
8k^{2}+8m_{s}^{2}}{\left( k^{2}-m_{s}^{2}\right) \left(
(k+p)^{2}-m_{s}^{2}\right) };$

\medskip \noindent $I_{2}^{P}(p)+I_{2}^{NP}(p) \equiv -\frac{1}{2}g^{2}\int
\frac{dk^{4}}{\left( 2\pi \right) ^{4}}\frac{8k^{2}+8m_{f}^{2}}{\left(
k^{2}-m_{f}^{2}\right) \left( (k+p)^{2}-m_{f}^{2}\right) }-\frac{1}{2}%
g^{2}\int \frac{dk^{4}}{\left( 2\pi \right) ^{4}}\cos (2p\widetilde{k})\frac{%
8k^{2}+8m_{f}^{2}}{\left( k^{2}-m_{f}^{2}\right) \left(
(k+p)^{2}-m_{f}^{2}\right) };$

\medskip \noindent $I_{3}^{P}(p)+I_{3}^{NP}(p) \equiv -\frac{1}{2}g^{2}\int
\frac{dk^{4}}{\left( 2\pi \right) ^{4}}\frac{8p{{\cdot} }k}{\left(
k^{2}-m_{f}^{2}\right) \left( (k+p)^{2}-m_{f}^{2}\right) }.-\frac{1}{2}%
g^{2}\int \frac{dk^{4}}{\left( 2\pi \right) ^{4}}\cos (2p\widetilde{k})\frac{%
8p{{\cdot} }k}{\left( k^{2}-m_{f}^{2}\right) \left(
(k+p)^{2}-m_{f}^{2}\right) };$

\medskip The planar terms involve integrations which are already done
ordinarily in field theory in commutative spacetime. Their contributions
lead, as in the commutative case, to logarithmic mass and wavefunction
renormalization. We are here mainly interested
in $\Sigma(p)_{1loop}^{NP(E)} $, the sum of the nonplanar contributions,
which we study in the euclidean region.
One easily finds$\footnote{$K_{0}(x)$ and $K_{1}(x)$
are modified Bessel functions of the second kind.}$
\begin{equation}
\Sigma _{1loop}^{NP(E)}(p)=I_{1E}^{NP}(p)+I_{2E}^{NP}(p)+I_{3E}^{NP}(p)~,
\label{revgamma}
\end{equation}
where
\begin{align}
I_{1E}^{NP}(p)& =\frac{g^{2}}{2\left( 2\pi \right) ^{2}}\int_{0}^{1}da\left%
\{ \left[ 8m_{s}^{2}+4p^{2}(1-a)(2a-1)\right] K_{0}(2\left| \widetilde{p}%
\right| \sqrt{m_{s}^{2}+p^{2}a(1-a)})+\right.  \nonumber \\
& \left. -\frac{4}{\left| \widetilde{p}\right| }\sqrt{m_{s}^{2}+p^{2}a(1-a)}%
K_{1}(2\left| \widetilde{p}\right| \sqrt{m_{s}^{2}+p^{2}a(1-a)})\right\} ,
\end{align}
\bigskip
\begin{equation}
I_{2E}^{NP}(p)=-\left[ I_{1E}^{NP}(p)\right] _{m_{s}\rightarrow m_{f}},
\end{equation}
\bigskip
\begin{equation}
I_{3E}^{NP}(p)=-\frac{4}{\left( 2\pi \right) ^{2}}p^{2}\frac{g^{2}}{2}%
\int_{0}^{1}dbbK_{0}(2\left| \widetilde{p}\right| \sqrt{m_{f}^{2}+p^{2}b(1-b)%
})~.  \label{cont3}
\end{equation}

In the case of exact SUSY, $m_{s}=m_{f}$, the contributions $I_{1E}^{NP}$
and $I_{2E}^{NP}$ cancel each other, so that $\Sigma
_{1loop}^{NP(E)}=I_{3E}^{NP}$ and there are no IR divergencies~\cite
{giro,MST}.

In the general case, $m_{s}\neq m_{f}$, IR divergencies are present. Their
structure depends on the relative magnitude of the SUSY-restoration scale $%
\Lambda_{SUSY}\simeq m_{f}$ and the noncommutativity scale $M_{nc}=\tfrac{1}{%
\sqrt{\left| \theta \right| }}$ (where $\left| \theta \right| $ denotes
generically a characteristic size of the elements of the matrix $\theta_{\mu
\nu}$).

If $M_{nc} < m_f$ and $p\ll \frac{M_{nc}^{2}}{m_f}$
the non-planar part of the self
energy is well approximated by
\begin{eqnarray}
\Sigma _{1loop}^{NP(E)}(p) &\simeq &\frac{g^{2}}{\left( 2\pi \right) ^{2}}%
\int_{0}^{1}da\left\{ 6m_{f}^{2}\ln \left( 2\left| \widetilde{p}\right|
\sqrt{m_{f}^{2}+p^{2}a(1-a)}\right) +\right.  \nonumber \\
&&-6m_{s}^{2}\ln \left( 2\left| \widetilde{p}\right| \sqrt{%
m_{s}^{2}+p^{2}a(1-a)}\right) +  \nonumber \\
&&+2p^{2}(1-a)(3a-1)\left[ \ln \left( \sqrt{m_{f}^{2}+p^{2}a(1-a)}\right)
-\ln \left( \sqrt{m_{s}^{2}+p^{2}a(1-a)}\right) \right]  \nonumber \\
&&+\left( m_{s}^{2}-m_{f}^{2}\right) [6\ln 2-6\gamma +1]+  \nonumber \\
&&\left. +2p^{2}a\left[ \ln \left( 2\left| \widetilde{p}\right| \sqrt{%
m_{f}^{2}+p^{2}a(1-a)}\right) -(\ln 2-\gamma )\right] \right\} ~.
\label{GIANLUCA1}
\end{eqnarray}
[This approximation is also valid for all $p < M_{nc}$
if $M_{nc} > m_f$, but we are mainly interested here
in the case $M_{nc} < m_f$ which allows us to explore the implications
for low-energy phenomena
of SUSY restoration above $M_{nc}$.]

If $M_{nc} < m_f$ and $\frac{M_{nc}^{2}}{mf}\ll p\ll M_{nc}$
the non-planar part of the self energy is well approximated by
\begin{align}
\Sigma _{1loop}^{NP(E)}(p)& \simeq \frac{g^{2}}{\left( 2\pi \right) ^{2}}%
\int_{0}^{1}da\left\{ -\frac{1}{\left| \widetilde{p}\right| ^{2}}+\right.
\nonumber \\
& -\ln \left( 2\left| \widetilde{p}\right|
\sqrt{m_{f}^{2}+p^{2}a(1-a)} \right)
\left[ 6m_{s}^{2}+2p^{2}(1-a)(3a-1)\right] +  \nonumber \\
& \left. +m_{s}^{2}[6\ln 2-6\gamma +1]+2p^{2}(1-a)[a(3\ln 2-3\gamma
+\frac{1}{2})-(\ln 2-\gamma )]\right\} .  \label{GIANLUCA2}
\end{align}

As a result of contributions coming from the UV portion of loop integrals,
we are finding that (for $m_{s}\neq m_{f}$) the model is affected by
logarythmic IR singularities (\ref{GIANLUCA1}) if $\frac{M_{nc}^{2}}{mf}\gg
p $, but as soon as momenta are greater than $\frac{M_{nc}^{2}}{m_{f}}$ the
dependence of the self-energy on momentum turns into an inverse-square law (%
\ref{GIANLUCA2}). In the limit $m_{f}\rightarrow \infty $, the case in which
there is absolutely no SUSY (not even in the UV), the inverse-square law
takes over immediately and the theory is affected by quadratic IR
singularities. The case of exact SUSY $m_{f}=m_{s}$ is free from IR
singularities, but of no interest for physics (Nature clearly does not enjoy
exact SUSY).

The IR/UV mixing manifests in two (obviously connected) ways which is worth
distinguishing: (1) The UV portion of loop integrals is responsible for some
IR singularities of the self-energies, (2) the low-energy structure of the
model can depend on $m_{f}$ even when $m_{f}$ is much higher than the energy
scales being probed. There is no IR/UV decoupling.

\subsection{Further effects on the low-energy sector from UV physics}
The implications of supersymmetry for the IR sector of canonical
noncommutative spacetimes are very profound. In our illustrative model one
finds that exact SUSY leads to absence of IR divergences, if SUSY is only
present in the UV (UV restoration of SUSY) one finds soft, logarythmic, IR
divergences, and total absence of SUSY ($m_{f} \rightarrow \infty$) leads to
quadratic IR divergences. While the presence of SUSY in the UV is clearly an
example of UV physics with particularly significant implications for the IR
sector of canonical noncommutative spacetimes, from this example we must
deduce that in general the loss of decoupling between UV and IR sectors can
be very severe. Other features of the UV sector, which perhaps have not even
yet been contemplated in the literature, might have similarly pervasive
implications for the IR sector.

A particularly interesting scenario is the one in which supersymmetry is
restored at some high scale (which in our illustrative model is $m_{f}$) and
then at some even higher scale, possibly identified with the so-called
``quantum-gravity scale", the theory predicts additional structures, which
in turn, again, would affect the infrared. The example of quantum gravity is
particularly significant since we have no robust (experimentally supported)
information on this realm of physics, so it represents an example of UV
physics for which our intuition migh easily fail, and as a consequence our
intuition for its implications for the IR sector of a field theory in
canonical noncommutative spacetime might also easily fail.

As a way to emphasize the sensitivity of the IR sector to such unknown UV
physics, it is worth noting here some formulas that describe features of our
illustrative model from the perspective of a theory with fixed cutoff scale $%
\Lambda$. For renormalizable field theories in commutative spacetime the
presence of such a cutoff would be basically irrelevant: if the cutoff is
much higher than all scales of interest it will negligibly affect all
predictions and it can be uneventfully removed through the limit $\Lambda
\rightarrow \infty $. Importantly, in a renormalizable field theory in
commutative spacetime the limit $\Lambda \rightarrow \infty$ is uneventful
independently of whether or not we have introduced in the theory all the
correct UV degrees of freedom hosted by Nature: the low-energy physics is
anyway independent of (decoupled from) the UV sector.

For field theories in canonical noncommutative spacetime the limit $\Lambda
\rightarrow \infty$ is not at all trivial, meaning that the
structures/degrees of freedom encountered along the limiting procedure can
in principle affect significantly the low-energy physics. One can take the $%
\Lambda \rightarrow \infty$ limit in a physically meaningful way only under
the assumption that one has complete knowledge of the full theory of Nature
(something which of course we cannot even contemplate).

The sensitivity of the IR sector to unknown UV physics is well characterized
by considering, for fixed cutoff scale $\Lambda $, the nonplanar
contributions to the two point functions. For the two-point function we
already considered previously one finds:
\begin{align}
I_{1E}^{NP}& =\frac{g^{2}}{2}\left\{ \frac{1}{\left( 2\pi \right) ^{2}}%
\int_{0}^{1}da\left[ 8m_{s}^{2}+4p^{2}(2a-1)(1-a)\right] K_{0}(2\sqrt{%
\widetilde{p}^{2}+\frac{1}{\Lambda ^{2}}}\sqrt{m_{s}^{2}+p^{2}a(1-a)}%
)+\right.  \nonumber \\
& \left. +\frac{4}{\sqrt{\widetilde{p}^{2}+\frac{1}{\Lambda ^{2}}}}\left[
\frac{\widetilde{p}^{2}}{\widetilde{p}^{2}+\frac{1}{\Lambda ^{2}}}-2\right]
\sqrt{m_{s}^{2}+p^{2}a(1-a)}K_{1}(2\sqrt{\widetilde{p}^{2}+\frac{1}{\Lambda
^{2}}}\sqrt{m_{s}^{2}+p^{2}a(1-a)})\right\}
\end{align}
\begin{equation}
I_{2E}^{NP}=-I_{1E}^{NP}(m_{s}\rightarrow m_{f})
\end{equation}
\begin{equation}
I_{3E}^{NP}=-\frac{4}{\left( 2\pi \right) ^{2}}p^{2}\frac{g^{2}}{2}%
\int_{0}^{1}dbbK_{0}(2\sqrt{\widetilde{p}^{2}+\frac{1}{\Lambda ^{2}}}\sqrt{%
m_{f}^{2}+p^{2}b(1-b)})
\end{equation}
Note that nonplanar diagrams are cutoff
by $\Lambda _{eff}=\frac{1}{\sqrt{\widetilde{p}^{2}+\frac{1}{\Lambda ^{2}}}}$.
The self-energy is insensitive to the value
of $\Lambda $ as long as the
condition $\left| \widetilde{p} \right| \gg \frac{1}{\Lambda }$ is satisfied.
But for $\left| \widetilde{p} \right| <\frac{1}{\Lambda }$ there is an
explicit dependence\footnote{It is worth noticing that for fixed
cutoff $\Lambda $ and $\left| \widetilde{p}\right| <\frac{1}{\Lambda }$
the self-energy is essentially independent of
the noncommutativity parameters. This is due to the fact that under those
conditions the nonplanar contributions are completely negligible.
This might encourage one to contemplate the possibility of a
physical cutoff scale $\Lambda $, but it is important to notice
that such a scale would be observer
dependent since ordinary Lorentz transformations still govern the
transformations between inertial observers in canonical noncommutative
spacetime~\cite{gacAREAnew}. (In other noncommutative spacetimes, where the
action of boosts is deformed, a cutoff scale can be introduced in an
observer-independent way~\cite{dsr1,gacAREAnew}, but this is not the case of
canonical noncommutative spacetimes.) We shall disregard this possibility;
however, in theories that already identify a preferred class of inertial
observers, such as theories in canonical noncommutative spacetimes, the
possibility of an observer-dependent cutoff scale cannot~ \cite{gacAREAnew}
be automatically dismissed.} on $\Lambda $ signaling that the infrared
sector is sensitive to new physics in the UV.

\section{Conditional bounds on noncommutativity parameters from low-energy
data}
The main point of our manuscript is that the observations made in the
previous Section have significant implications for the comparison of
low-energy experimental data with a theory in canonical noncommutative
spacetime.

It is useful to note here a brief description of the conventional technique
that allows to use low-energy data to set absolute (unconditional!) limits
on the parameters of theories in commutative spacetime:

\begin{itemize}
\item  {\bf 1C.} Data are taken in experiments involving particles with
energies/momenta from some lower (IR) limit, ${\cal S}_{min}$ (we of course
do not have available probes with wavelength, {\it e.g.}, larger than the
size of the Universe) up to an upper limit, ${\cal S}_{max}$, which
naturally coincides with the highest energy scales attainable in our
laboratory experiments (and, in appropriate cases, the energy scales
involved in certain observations in astrophysics).

\item  {\bf 2C.} We then compare these experimental results obtained at
energy/momentum scales within the range $\{{\cal S}_{min},{\cal S}_{max}\}$
to the corresponding predictions of the theory of interest. In deriving
these predictions we sometimes formally appear to use the whole structure of
the theory, all the way to infinite energy/momentum; however, in reality,
because of the IR/UV decoupling that holds in (renormalizable) theories in
commutative Minkowski spacetime, the theoretical prediction only depends on
the IR structure of the theory, up to energy/momentum scales which are not
much bigger than ${\cal S}_{max}$. (For example, degrees of freedom with
masses of order, say, $10^{5}{\cal S}_{max}$ would anyway not affect the
relevant predictions).

\item  {\bf 3C.} If the theoretical predictions obtained in this way do not
agree with the observations performed in the range $\{{\cal S}_{min},{\cal S}%
_{max}\}$ we then conclude that the theory in question is to be abandoned.

\item  {\bf 4C.} If the theoretical predictions obtained in this way agree
with the observations performed in the range $\{{\cal S}_{min},{\cal S}%
_{max}\}$ we then conclude that the theory in question provides a valid
description of phenomena up to energy/momentum scales of order ${\cal S}%
_{max}$. Typically the predictions of the theory will depend on some free
parameters and this parameter space will be constrained by the requirement
of agreeing with the observations. Values of the parameters that do not
belong to this allowed portion of the parameter space are definitely
(unconditionally) excluded, since nothing that we could introduce in the
ultraviolet could modify the low-energy predictions. In light of the fact
that the structure of the theory above ${\cal S}_{max}$ did not play any
true role in the derivation of the predictions, the successful comparison
with $\{{\cal S}_{min},{\cal S}_{max}\}$ experiments provides no particular
encouragement for what concerns the validity of the theory at scales much
above ${\cal S}_{max}$.

\item  {\bf 5C.} With precision measurements in the range $\{{\cal S}_{min},%
{\cal S}_{max}\}$ we can sometimes put limits on features of the theory also
slightly (up to a few orders of magnitude) above ${\cal S}_{max}$. For
example, one of the parameters of the theory could be the mass of a certain
particle and the contributions to low-energy processes due to that particle,
while suppressed by its mass, can be tested in high-precision measurements.
\end{itemize}

For theories in canonical noncommutative spacetime the situation is quite
different, as one infers from the analysis reported in the previous Section.
The comparison between the theory and data taken in the range $\{{\cal S}%
_{min},{\cal S}_{max}\}$ is much more delicate:

\begin{itemize}
\item  {\bf 2NC.} From the observations made in the previous Section it
follows that in a canonical noncommutative spacetime a truly reliable
derivation of the predictions for the energy/momentum range $\{{\cal S}%
_{min},{\cal S}_{max}\}$ requires full knowledge of the theory at all
energy/momentum scales up to $M_{nc}^{2}/{\cal S}_{min}$ (and of course, if $%
M_{nc}\gg {\cal S}_{max}$, the scale $M_{nc}^{2}/{\cal S}_{min}$ can be much
higher than both $M_{nc}$ and ${\cal S}_{max}$). In particular, the IR/UV
mixing is such that degrees of freedom with masses that are much above $%
{\cal S}_{max}$ still affect significantly the predictions of the theory in
the range $\{{\cal S}_{min},{\cal S}_{max}\}$.

\item  {\bf 3NC.} So the theory can only be taken as a full description of
Nature. It cannot be intended to give the right predictions only in some
low-energy limit. If the predictions of such a theory are found to be in
conflict with observations, it might still well be that the theory contains
the right low-energy degrees of freedom, and that the disagreement is due to
having adopted the wrong UV sector. So, from our more conventional
perspective (in which we try to identify theories that contain the right
degrees of freedom up to a certain scale) disagreement with observations
does not force us to abandon the theory: it only invites us to introduce
appropriate new physics in the UV sector.

\item  {\bf 4NC.} Similarly, if the theoretical predictions are found to
agree with the observations performed in the range $\{{\cal S}_{min},{\cal S}%
_{max}\}$ when some free parameters fall within a certain allowed portion of
parameter space, values of the parameters that do not belong to that region
of the parameter space cannot be conclusively excluded. They are excluded
only {\bf conditionally}, in the sense that their exclusion is only
tentative, pending further exploration of the UV sector. Think for example
of the illustrative model we considered in the preceding Section. The $%
m_{f}\rightarrow \infty $ of that model is a model without any SUSY (not
even in the UV sector). One could propose such a non-SUSY model and compare
it to data obtained in the range $\{{\cal S}_{min},{\cal S}_{max}\}$.
Clearly the need to agree with observations would then impose a severe
(lower) bound on the noncommutativity scale, a key parameter of the theory,
in order to suppress the IR divergences ({\it e.g.} effectively relegating
those divergences at scales below ${\cal S}_{min}$). However, this bound on
the noncommutativity scale would be only conditional, in the sense that
modifying the theory only in the ultraviolet ({\it i.e.} where we would say
it has not been tested with our data in the range $\{{\cal S}_{min},{\cal S}%
_{max}\}$) may be sufficient to lift the bound. In fact, SUSY in the
ultraviolet sector ($m_{f}$ large but finite) significantly softens the
divergences used to set the bound. Whereas in commutative spacetime the
bounds on parameter space apply directly to the structure of the theory in
the range of energy/momentum scales that have been probed experimentally, in
canonical noncommutative spacetime the information gained experimentally in
the range $\{{\cal S}_{min},{\cal S}_{max}\}$ leaves open two possibilities:
it may still, as in the case of theories in commutative spacetime, constrain
the parameters of the theory in that same range of energy/momentum scales,
but one cannot exclude the possibility that our low-energy observations are
instead primarily a manifestation of some features of the UV sector
(transferred to the low-energy sector via the IR/UV mixing) and therefore
cannot be used to constrain the low-energy structure of the theory. If there
is disagreement between theory and experiments in the range $\{{\cal S}%
_{min},{\cal S}_{max}\}$ one would normally assume that some aspects ({\it %
e.g.} the field content) of the theory must be changed in that same range of
energy/momentum scales, instead in canonical noncommutative spacetime that
same disagreement could be solved not only by introducing new features in
the $\{{\cal S}_{min},{\cal S}_{max}\}$ region but also by introducing new
features in the UV sector of the theory.

\item  {\bf 5NC.} Since data taken in the range $\{{\cal S}_{min},{\cal S}%
_{max}\}$ do not even give definitive information on the structure of the
theory in that same range, it is of course true that measurements in the
range $\{{\cal S}_{min},{\cal S}_{max}\}$ cannot be used to put limits on
features of the theory even just slightly above ${\cal S}_{max}$, no matter
how precise those measurements are. However, just because features of the UV
sector affect the low-energy physics, under the assumption that the
spacetime is indeed canonically noncommutative, one can gain insight of the
UV structure of the theory, even just using low-energy data. For example,
some of the observations made in the previous Section provide an opportunity
to discover UV SUSY even just using low-energy data: if data allowed us to
identify an energy/momentum scale at which the self-energy changed its
qualitative dependence on momentum in the way described by comparison of
Eqs.~(\ref{GIANLUCA1}) and (\ref{GIANLUCA2}), we could then infer rather
robustly the presence of SUSY at high energies and (if the value of the
noncommutativity scale was deduced from some other observations) we could
even deduce the scale of SUSY restoration.
\end{itemize}

\section{Futility of approaches based on expansion in powers of $\protect%
\theta$}
The observations reported in the preceding section indicate that some of the
standard techiques used in phenomenology require a prudent implementation in
the context of theories in canonical noncommutative spacetimes. We want to
emphasize in this section that for one of the techniques which served us
well in the analysis of theories in commutative spacetime there are even
more severe limitations to the applicability in the context of theories in
canonical noncommutative spacetimes. This is the technique that relies on
the truncation of a power series in one of the parameters of the theory: we
argue that, at the quantum-field-theory level, the results obtained by
truncating a power series in $\theta $ do not provide a reliable
approximation of the full theory. This type of truncation, which has been
widely used in the literature~\cite{GW,dh,ajsw,jmssw,H,cclnuova,BDDSTW}), is
based on the inclusion of only a few terms in the $\theta $-expansion of the
Moyal $\star$-product. For example up to the second order in $\theta$ one
could write
\begin{eqnarray}
\varphi _{1}(x)\star \varphi _{2}(x) &=&\varphi _{1}(x)\varphi _{2}(x)+\frac{%
i}{2}\theta ^{\mu \nu }\partial _{\mu }\varphi _{1}(x)\partial _{\nu
}\varphi _{2}(x)+  \nonumber \\
&&-\frac{1}{8}\theta ^{\alpha \beta }\theta ^{\mu \nu }\partial _{\alpha
}\partial _{\mu }\varphi _{1}(x)\partial _{\beta }\partial _{\nu }\varphi
_{2}(x)+O(\theta ^{3})  \label{expa}
\end{eqnarray}
The resulting action constructed with the truncated $\star $-product (\ref
{expa}) depends only on a finite number of derivatives so it is local,
unlike the full theory. Moreover, since $\theta $ has negative mass
dimensions, the action will also certainly be power-counting
nonrenormalizable, whereas the full theory might be renormalizable \cite
{mrs,giro,ABK,MJ,gri}.

Even more serious concerns emerge from the realization that the expansion
one is performing is (of course) not truly based on a power series in the
dimensionful quantity $\theta$: it is rather an expansion in dimensionless
quantities of the type $p\theta p$. Therefore already at tree level the
truncated $\theta$-expanded theory can only give a good approximation of the
full theory at scales $p$ such that $p\theta p\lesssim 1$, {\it i.e.} $%
p\lesssim 1/\sqrt{\theta }$.

But actually even in that range of momenta the expansion cannot be used
reliably. Its reliability is spoiled by quantum corrections. The quantum
corrections involve the Moyal $\star$-product inserted in loop diagrams, and
the truncation will reliably describe these loop corrections only for loop
momenta such that $p\lesssim 1/(\theta \Lambda )$. In fact, in loop
integrals involving factors of the type $p\theta k$, with $p$ playing the
role of external momentum and $k$ playing the role of integration/loop
momentum, one would like a reliable truncation that is valid over the whole
loop-integration range, which extends at least up to a cutoff $\Lambda$. In
order to have $p\theta k\lesssim 1$ even for $k$ as large as $\Lambda$ it is
necessary to assume that indeed $p\lesssim 1/(\theta \Lambda )$. This can
also be inferred straightforwardly in the illustrative example of the ``$%
\lambda \Phi ^{4}$'' scalar-boson field theory: there one finds that the
full theory predicts nonplanar terms giving a leading contribution of the
form
\begin{equation}
\Sigma _{NP}^{1}(p)\simeq \dfrac{g^{2}}{\widetilde{p}^{2}+1/\Lambda ^{2}}%
=\Lambda ^{2}\dfrac{g^{2}}{\Lambda ^{2}\widetilde{p}^{2}+1}.
\end{equation}
whereas the truncated $\theta $-expansion of the $\star $-product would
replace this prediction with
\begin{equation}
\Sigma _{NP}^{1}(p)\simeq g^{2}\Lambda ^{2}\left\{ 1-\Lambda ^{2}\widetilde{p%
}^{2}+O(\theta ^{4})\right\} .
\end{equation}
Clearly the two expressions are equivalent only if $\Lambda ^{2}\widetilde{p}%
^{2}\lesssim 1$, which indeed corresponds to $p\lesssim 1/(\theta \Lambda )$.

Therefore, when one includes quantum/loop effects,
the truncated $\theta$-expansion could be a good approximation
of the full theory only in the
range of momenta $p\lesssim 1/(\theta \Lambda )$. But as we have discussed
in the preceding section this is just the range of momenta in which the
theory is maximally sensitive to ultraviolet physics, which we must assume to
be unknown. In other words the truncated $\theta$-expansion reliably
approximates the full theory only in a regime where the full theory is
itself void of predictive power,
because of its sensitivity to unknown physics that might be present in
the ultraviolet. It therefore appears that these
truncated $\theta$-expansions cannot be used for
a meaningful comparison between data and
theories in canonical noncommutative spacetime.
In other contexts expansions in powers of $p$ versus some
characteristic momentum scale have been proven to give a reliable
low-energy effective-theory description of the full theory
one intends to study, but in this case of field theories in canonical
noncommutative spacetime the IR/UV mixing provides a powerful
obstruction for any attemp to obtain a meaningful
low-energy effective-theory description.

\section{Conclusions}
Clearly the type of IR/UV mixing which is present in field theories in
canonical noncommutative spacetime has wide implications for the strategies
that should be adopted in order to falsify/verify these theories. Theories
that (according to our conventional language) differ only in an
experimentally unaccessible range of momenta may give rise to different
predictions in the low-energy regime. The bounds on parameter space that one
usually is able to set using low-energy data are here only conditional, in
the sense clarified in Section~4. On the other hand low-energy data can be
used to gain insight on the UV sector, as we discussed for the specific case
of UV SUSY, under the assumption that the theory does indeed live in a
canonical noncommutative spacetime.

The implications of the IR/UV mixing are clearly very severe for
self-energies. In the models so far studied it appears~\cite
{mrs,MST,giro,ver} that instead the implications for interaction vertices
are less significant. This might mean that tests based on the properties of
interaction vertices could be more indicative (less conditioned on
assumptions concerning the UV sector) than tests based on properties of the
self-energies. However, it is perhaps best to be cautious in formulating
this expectation in general terms: in these theories we are not protected by
the usual IR/UV decoupling and the fact that in certain specific models one
finds that interaction vertices are only moderately sensitive to properties
of the UV sector cannot provide a general reassurance. 

Some of the points we raised here clarify that in the investigation of
theories in canonical noncommutative spacetime it might be necesssary to
``build a case'' in favour or against consistency with observations (whereas
in commutative spacetime a single experiment can give conclusive
unconditional indications). The case would be built by considering a variety
of data, and observing that they are all consistent with the characteristic
structure of theories in canonical noncommutative spacetime. Because of the
nature of these characteristic features of canonical noncommutativity, it
can be very useful to rely on data that concern a wide range of energy
scales. The astrophysical studies analyzed, for what concerns canonical
noncommutativity, in Ref.~\cite{gl} could play an important role in this
programme. For example, the larger range of energies explored, if
astrophysical data on particle propagation were combined with the
corresponding laboratory data, could allow to establish that at some
particular momentum scale the dependence of the self-energy on momentum
suffers a transition of the type described in Eqs.~(\ref{GIANLUCA1}) and (%
\ref{GIANLUCA2}), which is a characteristic feature of theories in canonical
noncommutative spacetime with (softly broken) UV SUSY.

\section*{Acknowledgments}
We thank V.V.~Khoze for conversations on this research topic during a
meeting in Rome (October 2001), where one of us (G. A.-C.) presented the
preliminary results of this analysis. It was reassuring for us to be
informed of the fact that Khoze and other canonical-noncommutativity experts
at the Univ. of Durham (U.K.) were reaching conclusions similar to ours. One
of us (G.M.) also acknowledges stimulating conversations with R.~Szabo.

\baselineskip12pt plus .5pt minus .5pt

\end{document}